\documentclass[aps,prl,reprint,groupedaddress]{revtex4-1}
\usepackage{graphicx}

\renewcommand{\vec}[1]{\mathbf{#1}}

\begin{document}


\title{Hyperuniform structures formed by shearing colloidal suspensions}


\author{Sam Wilken, Rodrigo E. Guerra, David J. Pine, and Paul M. Chaikin}
\affiliation{Center for Soft Matter Research, Department of Physics, New York University, New York 10003, USA}


\date{\today}

\begin{abstract}

In periodically sheared suspensions there is a dynamical phase transition characterized by a critical strain amplitude $\gamma_c$ between an absorbing state where particle trajectories are reversible and an active state where trajectories are chaotic and diffusive. 
Repulsive non-hydrodynamic interactions between ``colliding" particles' surfaces have been proposed as a source of this broken time reversal symmetry.
A simple toy model called Random Organization qualitatively reproduces the dynamical features of this transition. 
Random Organization and other absorbing state models exhibit hyperuniformity, a strong suppression of density fluctuations on long length-scales quantified by a structure factor $S(q \rightarrow 0) \sim q^\alpha$ with $\alpha > 0$, at criticality. 
Here we show experimentally that the particles in periodically sheared suspensions organize into structures with anisotropic short-range order but isotropic, long-range hyperuniform order when oscillatory shear amplitudes approach $\gamma_c$. 

\end{abstract}

\maketitle

At vanishingly small Reynolds numbers, the Navier-Stokes equations take the form of the time-reversible creeping-flow Stokes equation. 
Experiments aimed at exploring the time reversibility of Stokes flow in the presence of hard particles under cyclic shear reveal the surprising result that steady state particle trajectories at small strains $\gamma$ are reversible but at large strains are chaotic and diffusive \cite{pineChaosThresholdIrreversibility2005,corteRandomOrganizationPeriodically2008}. 
Subsequently it was understood that non-hydrodynamic interactions, ``collisions"\cite{corteRandomOrganizationPeriodically2008,metzgerIrreversibilityChaosRole2013}, lead to rearrangements of the particle configurations until the system evolves---self-organizes---into a configuration where there are no particle collisions---an absorbing state. 
Above a critical strain $\gamma_c$, however, the system can no longer find such a configuration and continues evolving indefinitely---an active state. 

A simple toy model, Random Organization (RO), captures much of the dynamics of this transition, predicting, for example, the divergence in the time to find an absorbing state below the transition or the time to reach steady state in the active phase as a function of $|\gamma - \gamma_c |$ \cite{corteRandomOrganizationPeriodically2008}. 
This RO model belongs to a class of absorbing state dynamical models that, because they exhibit greater activity in denser regions, tend to become hyperuniform at their critical points ~\cite{hexnerHyperuniformityCriticalAbsorbing2015,tjhungHyperuniformDensityFluctuations2015b,weijsEmergentHyperuniformityPeriodically2015}.
At criticality, their long length scale density fluctuations become vanishingly small~\cite{torquatoLocalDensityFluctuations2003} and are characterized by a structure factor $S(q \rightarrow 0) \sim q^{\alpha}$ where $\alpha > 0$.
The value of $\alpha$ is universal for a given dimension in this class of absorbing state transitions, e.g. $\alpha = 0.25$ in 3D \cite{hexnerHyperuniformityCriticalAbsorbing2015}.
Such random but hyperuniform particle distributions have been conjectured to have unique optical and scattering properties~\cite{florescuDesignerDisorderedMaterials2009, manIsotropicBandGaps2013, manPhotonicBandGap2013}.
They are impossible to achieve for equilibrium structures of particles with finite-range interactions.

\begin{figure}[b!]
\includegraphics[scale=1.05]{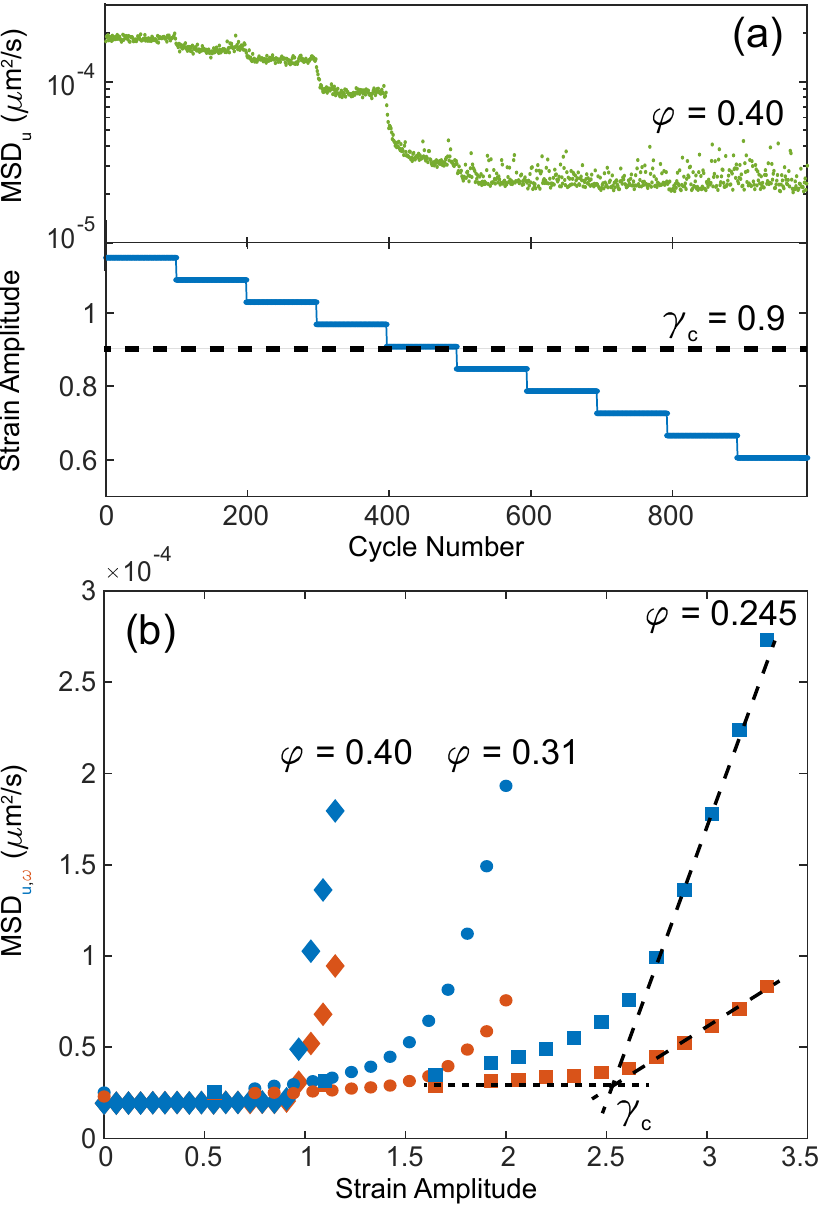}
\caption{\label{fig:msd} 
(a) The mean squared displacement (MSD), top panel, measures the magnitude of random displacements along the strain direction given the strain amplitude protocol in the bottom panel for a suspension of volume fraction $\phi$= 0.4. (b) The steady state MSD is plotted as a function of strain amplitude for both the strain (blue) and vorticity (orange) directions for suspensions of volume fraction $\phi$= 0.4, 0.31, and 0.245. }
\end{figure}
\begin{figure*}[t]
\includegraphics[width=16cm]{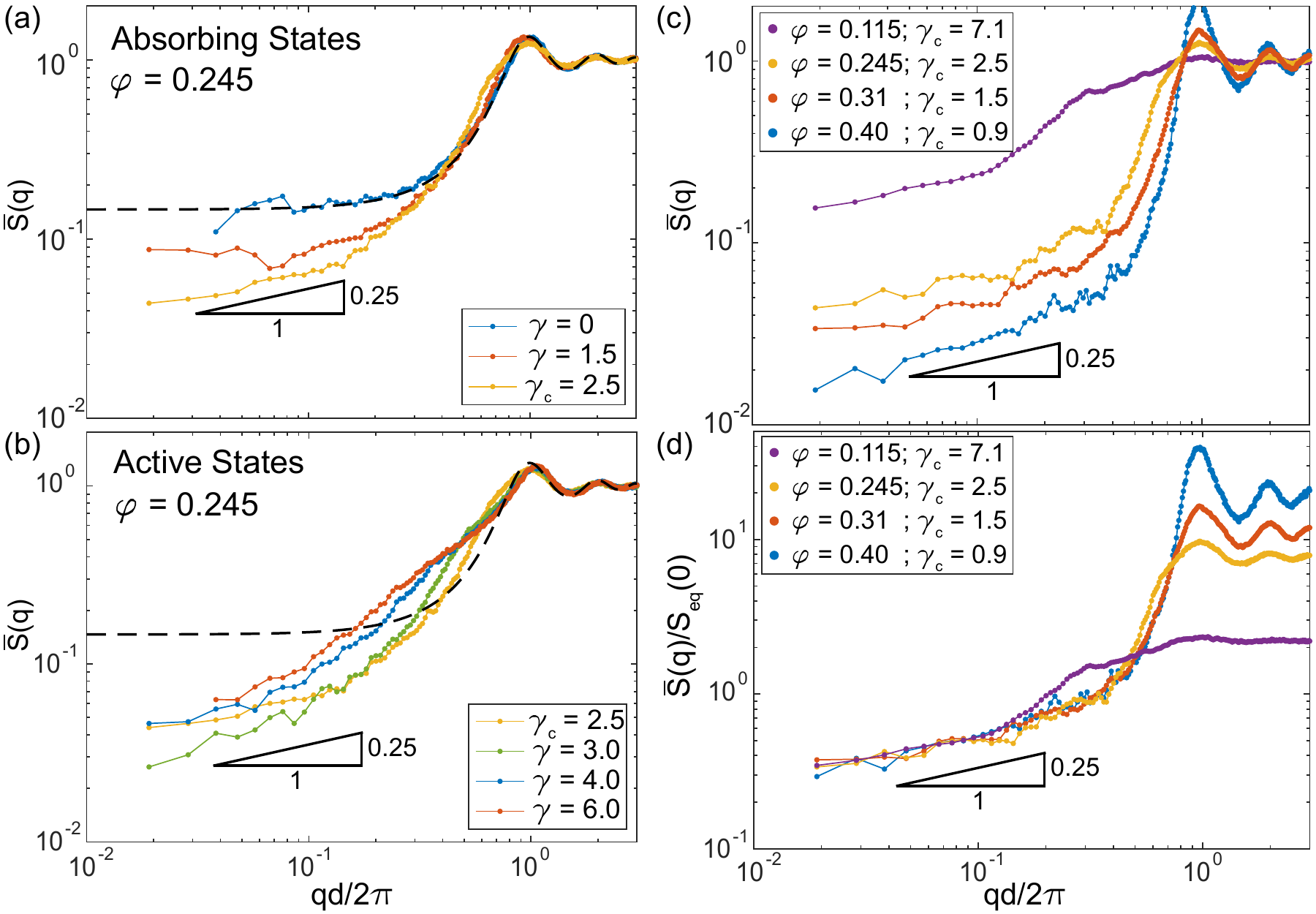}
\caption{\label{fig:Sqplot}
Structure factor of sheared suspension with volume fraction $\phi = 0.245$ for amplitudes (a) below and (b) above the critical value $\gamma_c = 2.5$ compared to equilibrium structure factor computed from Percus-Yevick approximation (dashed line). (c) Particle structure factors measured near the critical strain amplitude for volume fractions $\phi =$\,0.115, 0.245, 0.31, and 0.40. (d) Dividing measured $\bar{S}(q)$ shown in (c) by the equilibrium values of $S_{eq}(\phi, q\to 0)$ for each $\phi$ collapses all $\bar{S}(q)$ at small q. }
\end{figure*}
In this Letter we show experimentally that the transition between absorbing and active states of periodically sheared suspensions has at its critical point a structure that suppresses long-range density fluctuations and induces hyperuniform order in an otherwise fluid suspension of colloidal hard spheres.
While the quiescent suspension is isotropic, the shear strain breaks the isotropy and this is readily observed in $S(\vec{q} \sim 1)$.
Remarkably, $S(q)$ becomes isotropic as $q\to$\,0 with $S(q) \sim q^{0.25}$ when the periodic strain amplitude approaches $\gamma_c$.
We show that these dynamical and long-range structural features are reproduced quantitatively in simulations of an RO model that is modified to include a repulsive bias in the random displacements assigned to colliding particles. 
This minimal modification accounts for the increased excluded volume of suspensions at higher densities and provides evidence of universality for these dynamical and structural non-equilibrium phenomena.

Our colloid is composed of monodisperse $d = 1.20$-$\mu$m-diameter copolymer  particles~\cite{kodgerPreciseColloidsTunable2015} dispersed in 1-hexyl-3-propanenitrile imidazolium chloride~\cite{ziyadaDensitiesViscositiesRefractive2012}, a room-temperature ionic liquid with a viscosity of 20\,Pa$\cdot$s, 20,000 times the viscosity of water.
The chemical composition of the random copolymer, methyl methacrylate, trifluoroethyl methacrylate, and tert-butyl methacrylate, is chosen to match the refractive index and density of the particles to that of the suspending fluid and to allow surface grafting of a polyelectrolyte brush that provides short-range, effectively hard-sphere, repulsive interactions between particles.

We use a piezo-controlled rotation stage to precisely shear the suspension using a three degree cone-plate geometry.
We shear the sample periodically using a symmetric triangular waveform and employ a standard coordinate system: $\hat{u}$ is the flow direction, $\hat{\nabla}$ is the gradient direction, and $\hat{\omega}$ is the vorticity direction.
We image the particle suspension in three dimensions using a confocal microscope.
The spatial resolution is lower along the $\hat{\nabla}$ direction, as this corresponds to the confocal $z$-direction.

To identify the critical strain $\gamma_c$, we image a 2D horizontal slice in the $\hat{u}$-$\hat\omega$ plane once per shear cycle and track the mean squared displacement (MSD) \cite{crockerMethodsDigitalVideo1996} of particles as a function of applied strain, where the period of one cycle is eight seconds.
We  initialize the system using 100 cycles at $5 \gamma_c$ to erase any memory of previous runs.
We then use a step-wise decreasing strain ramp, starting at $\gamma > \gamma_c$ and strain for 100 cycles at each $\gamma$, which allows us to identify the transient and steady-state MSD per cycle.
As expected from previous studies, large strain amplitudes where $\gamma>\gamma_c$ produce large random displacements in steady state, while smaller strain amplitudes where $\gamma<\gamma_c$ produce no particle displacements beyond those of ordinary thermal Brownian motion. 
The Brownian motion of the particles is characterized by a diffusion coefficient $D = 2 \times 10^{-5}~\mathrm{\mu m^2/s}$ measured in a very dilute unsheared suspension.
The thermal diffusion is small, with typical displacements that are only 3\% of the particle diameter per cycle.
The Brownian diffusion provides a lower limit to the measured MSDs but the dynamical behavior separating the absorbing and active states is still clearly evident, as shown in Fig.~\ref{fig:msd}(a).

\begin{figure}
\includegraphics[scale=0.9]{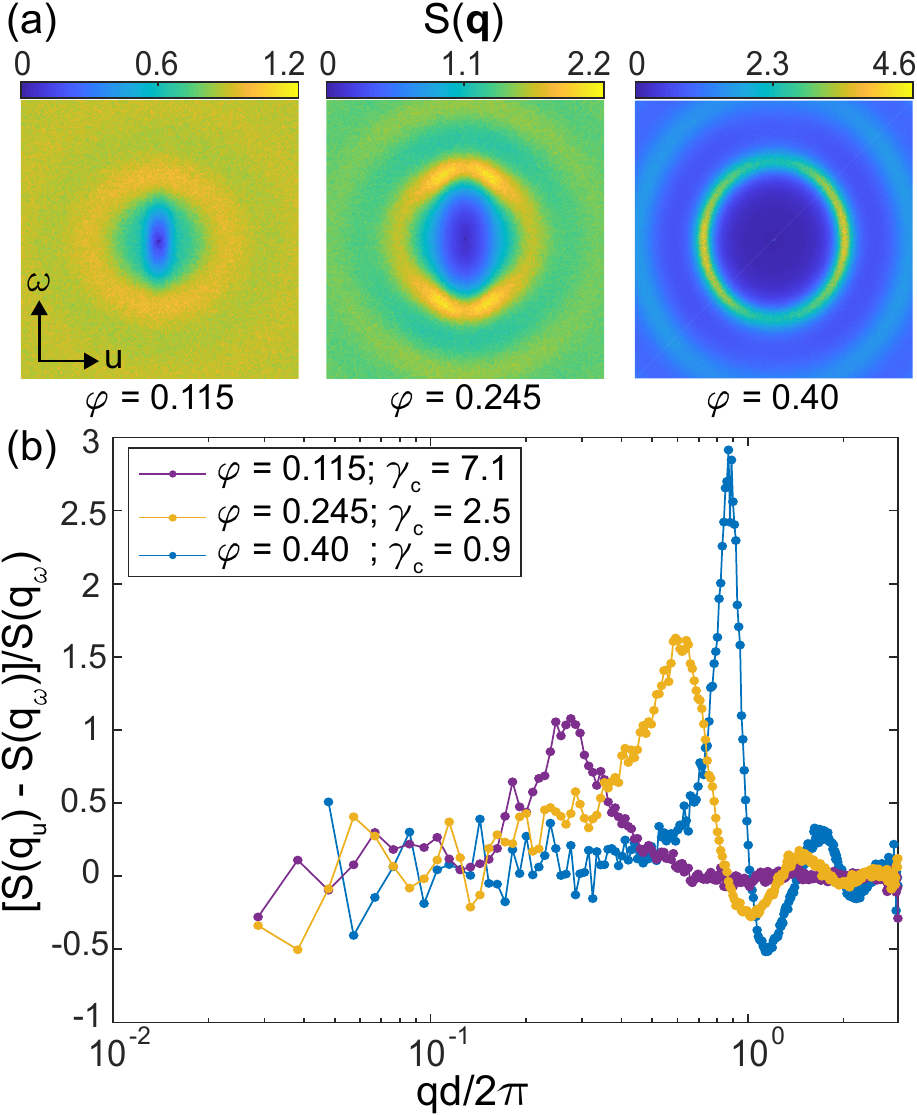}
\caption{\label{fig:Sqaniso}  
(a) Structural anisotropy is manifested in $S(q_u, q_\omega)$ by differences along two principle axes of shear, velocity and vorticity, for critically sheared suspensions with $\phi = 0.115$, 0.245, and 0.40. (b) The relative difference of $[S(q_u)- S(q_\omega)]/S(q_\omega)$ along two directions shows the strength of anisotropy as a function of $q$.}
\end{figure}
\begin{figure*}
\includegraphics[scale=0.95]{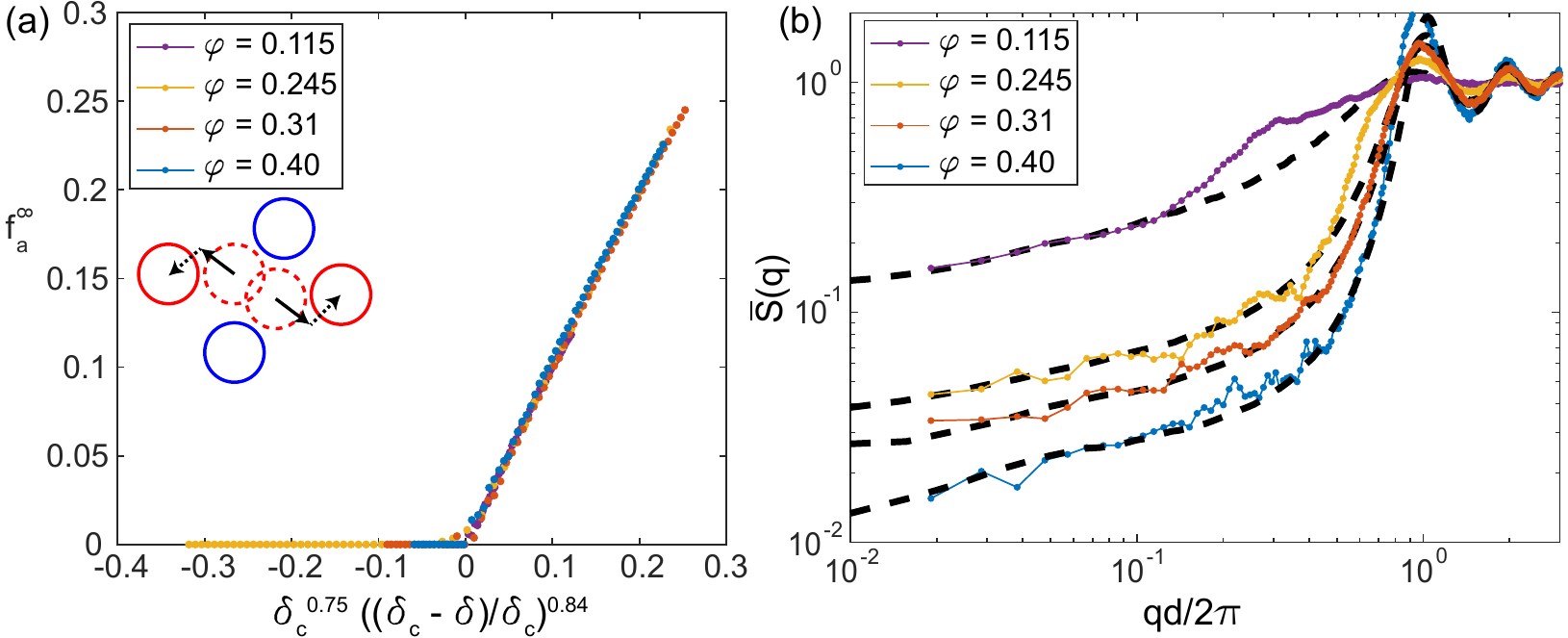}
\caption{\label{fig:brandorg} 
(a) The fraction of active/overlapping particles in steady state for the repulsive Random Organization model, $f_a^\infty$, plotted as a function of the reduced control parameter, $(\delta_c - \delta)/\delta_c$. 
Inset: Active particles (red) receive two displacements per cycle: a repulsive one (solid arrow) and a random isotropic one (dashed arrow), while isolated particles (blue) do not move.
(b) Structure factors of simulated (dashed lines) and experimental (points) configurations measured at $\delta_c(\phi)$ and $\gamma_c$ agree remarkably well with each other at small $q$ for every volume fraction.   }
\end{figure*}

We perform this ramp-down protocol on suspensions with volume fractions $\phi=0.245$, 0.31, and 0.40.
The transition to the absorbing state is rounded by Taylor dispersion at smaller volume fractions due to a larger strain rate and a slight increase in Brownian diffusion~\cite{taylorDispersionSolubleMatter1953, torrey_bloch_1956}.
Therefore, we identify $\gamma_c$ as the strain at which the linearly extrapolated MSD per cycle in the chaotic regime intersects the strain-independent thermal Brownian diffusion.
We find that the transition sharpens at higher volume fractions, with critical strains $\gamma_c$ that decrease with increasing $\phi$.
The inherent anisotropy of the applied shear is reflected in the much larger MSD of particles along $\hat{u}$ than along $\hat\omega$.
In more dilute samples, this anisotropy is also evident in the reversible state, as the residual Brownian motion along the flow direction is also amplified by Taylor dispersion, as shown in Fig.~\ref{fig:msd}(b).

To examine the effect of this dynamical transition on the structural properties of the suspension, we image a three dimensional volume of the suspension between shear cycles and compute the structure factor of the particle density, $S(\vec{q})=\langle\tilde{\rho}(\vec{q})\tilde{\rho}(-\vec{q})\rangle/V\langle\rho\rangle$, where $\tilde{\rho}(\vec{q})=\sum_ie^{-i\vec{q}\cdot \vec{x}_i}$, is the Fourier transform of the measured particle density.
Because we can only image 25$d$ into the sample, we integrate the particle density along $\hat{\nabla}$ and calculate $S(q_u,q_\omega,q_\nabla=0)$ with a non-equispaced FFT (NFFT)~\cite{keiner2009using}.
As shown in the SI, this two dimensional projection does not change the critical exponent $\alpha$.

For unsheared suspensions in thermal equilibrium, $S(\vec{q})$ of the suspension is well approximated by the Percus-Yevick structure factor for hard spheres~\cite{hansen_theory_2013}.
To measure the steady-state value of $S(\vec{q})$ for sheared suspensions we apply a large initial strain amplitude, $\gamma = 30$, to erase any memory of previous measurements and immediately decrease the strain amplitude to the prescribed $\gamma$ value, recording three-dimensional images and particle positions as the suspensions reach steady state.
From these transient measurements, we use exponential fits for every $q$-value to extrapolate the steady state $S(\vec{q})$. 
To reduce noise, we angularly average each $q$, denoted as $\bar{S}(q)$.

For $\gamma < \gamma_c$, $\bar{S}(q)$ does not differ significantly from its equilibrium behavior for length scales less than a few particle diameters, that is  for $q\gtrsim\pi/d$.
By contrast, for $q<\pi/d$, $\bar{S}(q)$ decreases continuously as $\gamma$ approaches $\gamma_c$, indicating a growth in long range correlations.
At $\gamma_c$, the system exhibits hyperuniform scaling: $\bar{S}(q,\gamma_c)\sim q^\alpha$, with $\alpha$= 0.25 $\pm$ 0.03, as shown in Fig.~\ref{fig:Sqplot}(a).
For $\gamma > \gamma_c$, $\bar{S}(q,\gamma_c)$ tends to increase for $q<\pi/d$, and its deviation from equilibrium becomes more pronounced, as shown in Fig.~\ref{fig:Sqplot}(b).

This critical scaling of the long-range density correlations of suspensions sheared at $\gamma_c$ is evident over a wide range of volume fractions, as shown for measurements of suspensions with $\phi=$\,0.115, 0.245, 0.31, and 0.40 shown in Fig.~\ref{fig:Sqplot}(c).
Moreover, we find that it is possible to collapse the small $q$ values of $\bar{S}(q,\gamma_c)$ for all of these volume fractions by scaling them by the Percus-Yevick  $S_{eq}(0)$, as shown in Fig.~\ref{fig:Sqplot}(d). 
All critically sheared suspensions fit the form $\bar{S}(q,\gamma_c)/S_{eq}(0) = A q^{0.25}$ at small $q$, where $A=1$.

This universal scaling observed at small $q$, however, is not seen at large $q$, where the anisotropy imposed by the shear flow is evident.
Angularly resolved plots of $S(\vec{q})$ show clear anisotropy for lower volume fractions, with correspondingly higher critical strain amplitudes.
But even dense suspensions show anisotropic correlations at length scales corresponding to the typical inter-particle spacing, as shown in Fig.~\ref{fig:Sqaniso}(a).
Nevertheless, the difference in $S(\vec{q})$ measured along the flow direction $S(q_u)$ and along the vorticity direction $S(q_\omega)$ vanishes at small $q$ and becomes isotropic for all measured volume fractions (Fig.~\ref{fig:Sqaniso}(b)).

The experimental data show clear dynamical and structural scalings near $\gamma_c$ that are qualitatively shared by the RO model, including the value of the scaling exponent, $\alpha\approx0.25$, but with a different dependence of $\gamma_c$ on $\phi$.
In the RO model, the critical strain $\gamma_c$ is large for small $\phi$ and goes to zero near $\phi=0.2$, meaning there are no absorbing states and no critical dynamics for $\phi>0.2$.
In experiments, however, we measure critical dynamics for volume fractions up to $\phi = 0.4$.
Therefore, we cannot quantitatively account for the experimental observations within the simple RO model.

To better account for our experimental observations, we modify the interactions in the model.
Between each simulation step, pairs of overlapping particles are given deterministic equal and opposite repulsive displacements of magnitude $\epsilon_d$ along the line connecting their centers.
In addition, they are given a displacement in a random direction of magnitude $\epsilon_r$ chosen from a Gaussian distribution of width $\epsilon_{r_0}$.
The relative magnitude of these displacements is specified by a control parameter $\delta = \epsilon_d/(\epsilon_d+\epsilon_{r_0})$, where $\delta = 0$ corresponds to simple RO.
Here we report results with $\epsilon_d+\epsilon_{r_0}=0.1d$; the results are insensitive to this choice.
The $\delta = 1$ limit corresponds to zero temperature Brownian dynamics with constant repulsive pairwise interactions, like the 3D monodisperse simulations in~\cite{ohernJammingZeroTemperature2003}, and we recover $\phi_c\to 0.639$, close to the random close packing volume fraction of monodisperse spheres. 
For simplicity we only consider $\gamma = 0$, so the control parameters for the transition are now $\phi$ and $\delta$ instead of $\phi$ and $\gamma$.
From the simulations we obtain the steady-state fraction of active particles $f_a^\infty$ as a function of the reduced control parameter ($\delta_c - \delta)/\delta_c$, and find that $f_a^\infty \sim ((\delta_c - \delta)/\delta_c)^{0.84}$ in the active phase, which is consistent with the Manna model critical exponent for activity $\beta = 0.84$~\cite{henkelNonequilibriumPhaseTransitions2008}. 
The data collapse onto a single curve when using a model dependent prefactor $\delta_c^{0.75}$, as shown in Fig.~\ref{fig:brandorg}(a).

We calculate the structure factor $S(q)$ at the critical control parameter $\delta_c(\phi)$ for each of the volume fractions used in the experiments.
The results of the simulations and the experiments show remarkable agreement at small $q$ with no adjustable parameters, as shown in Fig.~\ref{fig:brandorg}(b). 
This suggests that the modified RO model is in the same universality class as the simple RO model and other absorbing state models exhibiting the same critical exponent $\alpha$.
The deviations of $S(q)$ from the experimental measurements for larger $q$ stems from the lack of anisotropy in the model: smaller volume fractions exhibit larger deviations from experimental data for $q \gtrsim 0.2\pi/d$ because of their greater anisotropy, consistent with the data in Fig.~\ref{fig:Sqaniso}(b). 

The fact that experimental critical density correlations measured by $S(q)$ over a wide range of volume fractions exhibit the same long-range scaling behavior as an entire class of dynamical models and that a minimal modification to Random Organization is sufficient to quantitatively reproduce the small $q$ behavior of experimental data, suggest that the mechanism that drives self-organization in these critical dynamical systems towards hyperuniform structures is universal and robust.
Here, we see that it is tolerant of Brownian diffusion that smooths the transition between absorbing and active phases.

Although disordered hyperuniformity has not been found theoretically or experimentally for any equilibrium system with short range interactions, this paper demonstrates that hyperuniform materials are available through non-equilibrium processing. In this case, the result of an absorbing state transition. This opens the question of whether hyperuniform materials can be made by other dynamical methods and provides such materials for investigation of their potentially useful physical properties, such as band gaps for transmission of different types of waves.

\begin{acknowledgments}
This work was supported primarily by the MRSEC Program of the National Science Foundation under Award Number DMR-1420073 (SW) and partially supported by NASA under grant NNX13AR67G (REG) and the Center for Bio-Inspired Energy Sciences (CBES), an Energy Frontier Research Center funded by the U.S. DOE, Office of Sciences, Basic Energy Sciences under Award DE-SC0000989 (PMC, DJP).
\end{acknowledgments}

\end{document}